\documentclass[lettersize,journal]{IEEEtran}
\usepackage{amsmath,amsfonts,amssymb}
\usepackage{algorithmic}
\usepackage{array}
\usepackage[caption=false,font=normalsize,labelfont=sf,textfont=sf]{subfig}
\usepackage{textcomp}
\usepackage{stfloats}
\usepackage{url}
\usepackage{verbatim}
\usepackage{graphicx}
\usepackage[colorlinks = true, linkcolor = blue, citecolor = blue, urlcolor = blue]{hyperref}
\def\BibTeX{{\rm B\kern-.05em{\sc i\kern-.025em b}\kern-.08em
    T\kern-.1667em\lower.7ex\hbox{E}\kern-.125emX}}
\usepackage{balance}

\begin{document}

\title{Compact Superconducting Kinetic Inductance Traveling Wave Parametric Amplifiers with On-chip rf Components}
\author{L. Howe \IEEEmembership{}, A. Giachero, M. Vissers, J. Wheeler, J. Austermann, J. Hubmayr, J. Ullom
\thanks{L. Howe and J. Ullom are with the Quantum Sensors Division, National Institute of Standards and Technology, Boulder, CO 80305 USA; and with the Department of Physics, University of Colorado, Boulder, CO 80309 USA (e-mail: logan.howe@nist.gov)}
\thanks{A. Giachero is with the Quantum Sensors Division, National Institute of Standards and Technology, Boulder, CO 80305 USA; and with the Department of Physics, University of Colorado, Boulder, CO 80309 USA; and with the Department of Physics, University of Milano-Bicocca, 20126, Milan, Italy}
\thanks{M. Vissers, J. Wheeler, J. Austermann, and J. Hubmayr are with the Quantum Sensors Division, National Institute of Standards and Technology, Boulder, CO 80305 USA}
}


\maketitle

\begin{abstract}
Quantum computing systems and fundamental physics experiments using superconducting technologies frequently require signal amplification chains operating near the quantum limit of added noise. Both Josephson parametric amplifiers (JPAs) and traveling wave parametric amplifiers (TWPAs) have been used as first-stage amplifiers to enable readout chains operating within a few quanta or less of the quantum limit. These devices are also presently entering the commercial industry. However, nearly all demonstrations and existing products require bulky external microwave components for interconnection and application of requisite biases. These components -- cabling interconnects, bias tees, directional couplers, and diplexers -- increase the overall amplifier footprint, installation complexity, and reduce already limited available cryogenic volumes. Additionally, these components introduce loss and reflections which impact the measurement efficiency and readout system noise performance; thus making it more difficult to operate near the quantum limit.

Here we present the design and validation of microfabricated bias tees and directional couplers for operating three-wave mixing kinetic inductance TWPAs (KITs). We report the performance of KITs integrated with the microfabricated rf components. Using these devices we demonstrate reduction in the amplifier installation footprint by a factor of nearly five and elimination of all external, lossy microwave components previously required to operate a KIT. Our device displays a 2.8~GHz 3~dB bandwidth with a median true gain of 17.5~dB and median system noise of 3.4~quanta. These efforts represent the first full integration  of all rf components mandatory for TWPA operation on-chip. Our results mark significant progress towards the miniaturization and simplification of parametric amplifier setups and will aid in their more widespread applicability.

\end{abstract}

\begin{IEEEkeywords}
Traveling wave parametric amplifier, kinetic inductance, quantum-limit, compact, efficient, superconducting, quantum computing
\end{IEEEkeywords}

\section{Introduction}
The rapid growth in quantum information science and technology in recent years has spawned a large need for amplifiers operating near the quantum limit (QL) of added noise. Additionally, future rare physics searches \cite{liu2022bread} and cryogenic detector array readout \cite{malnou2023improved, dober2021microwave} can benefit from readout system noise levels approaching the QL. Many of these applications require these quantum-limited first-stage amplifiers (FSAs) to possess instantaneous bandwidths in excess of $\sim 2$~GHz, high dynamic range (above -80~dBm input power), the ability to operate in large magnetic fields ($\gtrsim 5$~T), and at high frequencies ($\sim 1$~THz) and temperatures ($\gtrsim 4$~K) \cite{malnou2022performance, liu2022bread}. Conventional Josephson parametric amplifiers (JPAs) satisfy very few of these criteria and, while Josephson traveling wave parametric amplifiers (JTWPAs) or impedance-matched JPAs are capable of increasing the gain bandwidth \cite{kaufman2023josephson, qiu2023broadband, mutus2014strong}, these amplifiers are still largely unsuitable in scenarios requiring moderate dynamic range, high temperature or high frequency operation, and ambient magnetic fields.

Conversely, amplifiers based on the nonlinear superconducting kinetic inductance are promising solutions to nearly all the limitations of Josephson-based parametric amplifiers. They have been demonstrated with multi-gigahertz bandwidths \cite{malnou2021three, shu2021nonlinearity, giachero2024kinetic}, input dynamic ranges as high as -63~dBm \cite{malnou2021three}, at elevated frequencies \cite{tan2024operation} and temperatures \cite{malnou2022performance}, and in multi-tesla ambient fields \cite{zapata2024granular, vaartjes2024strong}.

The main limitations in implementing kinetic inductance TWPAs (KITs) lie in the historically large pump powers required to achieve high gain (above -30~dBm at the KIT input), and in the suite of external (often commercial) microwave components required to supply the requisite bias and pump signals to the KIT. Recently an order of magnitude reduction in the required pump power was demonstrated via usage of a thinner NbTiN layer (reduced from 20~nm to 10~nm) \cite{giachero2024kinetic} -- making these devices more applicable to reading out ultra-sensitive devices such as superconducting qubits. These devices now require around -45~dBm of pump power at the KIT input.

\begin{figure*}
    \centering
    \includegraphics[width = .99\textwidth]{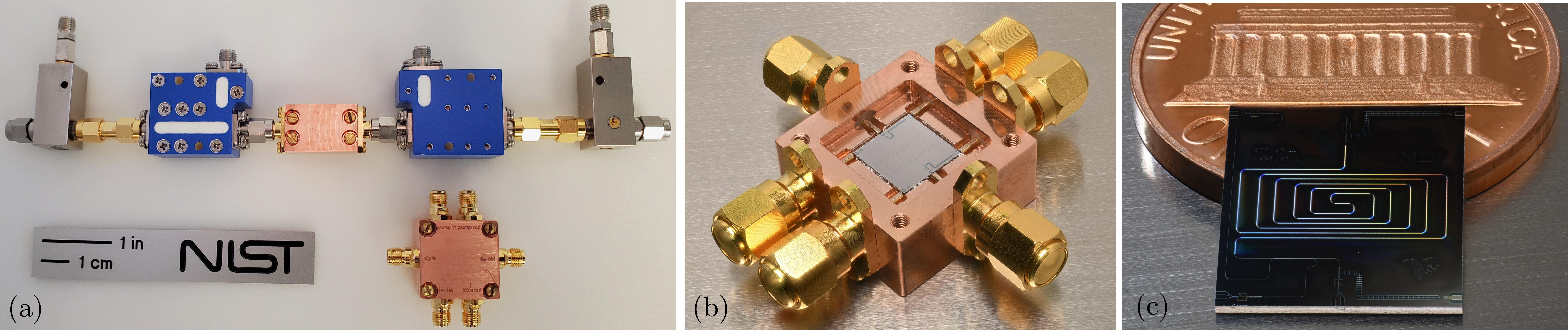}
    \caption{\textbf{(a)} Photograph of a typical minimal setup for integration of a KIT compared to the ORCK. The top row of components shows a traditional KIT with only a single input and output port (copper box in center) and a pair of diplexers (blue) and bias tees (silver). Below this we show the ORCK package and emphasize that no additional external components are \textit{required} to operate the ORCK itself -- however external filters or isolators may be necessary, e.g. in superconducting qubit readout, purely for the protection of the sensitive DUTs. \textbf{(b)} An ORCK in a wirebonded package that eliminates interfacial PCBs and the corresponding inevitable impedance discontinuities and reflections. Instead, we perform electromagnetic simulations to optimize the launch of an SMA connector with a protruding pin and bond directly to this pin. \textbf{(c)} Close-up of the ORCK on its \mbox{10~mm $\times$ 10~mm} die. The signal in (out) launches are on the bottom (top) center, while the pump in (out) are on the left bottom (top), and the bias in (out) are on the right bottom (top) launches. The $\sim8$~cm amplification medium (stub-loaded inverted microstrip transmission line) can be seen as the multicolored double spiral in the chip center.}
    \label{fig:kit_2port_vs_orck_picture}
\end{figure*}

However, the ultimate achievable readout system noise \cite{malnou2024low} and integration feasibility are hampered by the need for multiple centimeter-scale external rf components whose combined loss can quickly exceed 1~dB. Our KIT fabrication process natively uses very low loss dielectrics and superconducting metal films -- which are ideally suited to improve KIT installations via on-chip integration of requisite bias hardware with negligible loss. We call such an integrated device the on-chip rf component KIT (ORCK) \cite{howe2025integrated} and in this work we demonstrate its ability to compactify, simplify, and enhance readout chain noise performance. Indeed, Fig.~\ref{fig:kit_2port_vs_orck_picture}(a) demonstrates reduction in the overall KIT installation footprint by more than a factor of five with the ORCK. While some bias structures have been recently demonstrated on-chip \cite{nilsson2024superconducting} (in the form of the dc bias lines), we note that the ORCK is the first TWPA demonstrating on-chip integration of \textit{all} rf components required for its operation.

\section{Kinetic Inductance Traveling Wave Parametric Amplifiers}
Any electrical nonlinearity which can be modulated is capable of producing parametric interactions/amplification. In this way, Josephson-junction-based parametric amplifiers use the Josephson inductance to drive the parametric processes; we instead use the superconducting kinetic inductance in a thin layer of the disordered superconductor NbTiN. The kinetic inductance, $L_k$, as a function of the current, $I$, may be Taylor-expanded \cite{malnou2021three}
\begin{equation}
    L_k(I) = L_{\text{dc}} \left[ 1 + \left(\frac{I}{I_*}\right)^2 + \left(\frac{I}{I_*}\right)^4 + ... \right]
\end{equation}
and re-written as
\begin{equation}
    L_k(I) = L_{\text{dc}} [ 1 + \varepsilon I_p + \xi I_p^2 + \mathcal{O}(I^3)].
\end{equation}
$I_*$ sets the scale of the $L_k$ nonlinearity and here we separate the total current into the dc and pump currents as $I = I_{\text{dc}} + I_p$. The prefactors $\varepsilon$ and $\xi$ describe the relative strength of the three-wave-mixing and four-wave-mixing processes, respectively, as
\begin{equation}
    \varepsilon = \frac{2 I_{\text{dc}}}{I_*^2 + I_{\text{dc}}^2}~, ~~~ \xi = \frac{1}{I_*^2 + I_{\text{dc}}^2}.
\end{equation}
$I_*$ is proportional to the cross-sectional area of the high-$L_k$ transmission line wiring and is generally a few times higher than the line's critical current $I_c$. The NbTiN films used in this work with a 10~nm thickness and $1~\mu$m width have $L_{\text{dc}} = 35$~pH/$\square$, critical currents around $800~\mu$A, and $I_*$ of 2.8~mA \cite{giachero2023characterization}.

\subsection{Device Design}
The amplifiers discussed in this paper are three-wave-mixing amplifiers which are created from an 8.16~cm length of \textit{amplification medium}: a stub-loaded inverted microstrip NbTiN transmission line (1~$\mu$m width, 10~nm thickness) \cite{shu2021nonlinearity, klimovich2024investigating} with stepped dispersion engineering \cite{malnou2021three} to enforce broadband phase-matching. I.e. the amplification medium is composed of a periodic chain of unit cells with matched (\textit{unloaded}) and mismatched (\textit{loaded}) characteristic impedances. The combination of a segment of unloaded and loaded cells is called a \textit{supercell} and the number of each cell type within the supercell need not be equal. The configuration for the KITs in this work consists of 30 unloaded ($Z_0$ target of 50~$\Omega$) unit cells, and 4 loaded ($Z_0$ target of 80~$\Omega$) cells. Each unit cell is 2~$\mu$m in length and we place 1200 supercells in a double spiral geometry to ensure a minimum gain of over 20~dB (determined via simulation).

Prior KITs utilized a stub-loaded coplanar waveguide (CPW) transmission line topology with a narrow centerline (minimum width of $1~\mu$m limited by optical lithography resolution) to increase $L_k$ and decrease $I_*$ as much as possible for a fixed NbTiN film thickness -- making parametric amplification accessible at pump powers more practical for superconducting device readout \cite{malnou2021three, malnou2022performance, malnou2023improved}. These devices suffered from low yield due to persistent short-to-ground failures in the CPW centerline. Switching to an inverted microstrip topology \cite{shu2021nonlinearity, giachero2024kinetic} with a low-loss $\alpha$-Si dielectric (100~nm thickness) almost entirely eliminates this failure mode and -- as the phase velocity is $v_{ph} = 1/\sqrt{\mathcal{LC}}$ -- further miniaturizes the amplification medium by increasing the unit length capacitance $\mathcal{C}$. 

\section{\label{sec:noise_model}Noise Model}
The primary goal of any parametric amplifier installation is to lower the system noise. Via the Friis equation \cite{malnou2024low, pozar2021microwave}, and the axiom that superconducting parametric amplifiers are sufficiently low-loss as to fundamentally operate at the QL, a parametric power gain of $\sim 20$~dB is desirable. In this case the parametric amplifier, operating with an added noise near the QL, amplifies this 0.5~quanta of noise well above the added noise of the second-stage amplifier and thus sets the entire readout chain's noise to the QL. These second-stage amplifiers are typically semiconductor (HEMT) amplifiers operating with typical gains of $\gtrsim 35$~dB, at temperatures near 4~K, and adding noise of $\lesssim 10$~quanta \textit{at their input reference plane}. 

\begin{figure*}[t]
    \centering
    \includegraphics[width=0.9\textwidth]{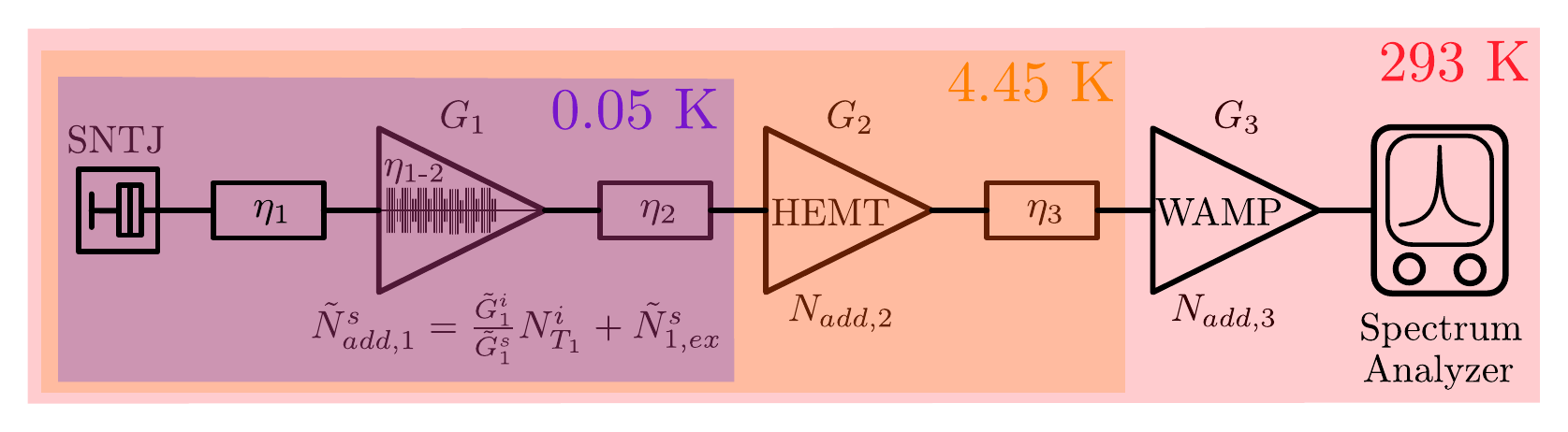}
    \caption{Schematic of the three-stage amplifier readout chain used to characterize the system noise, in a dilution refrigerator, while using the ORCK as the FSA. A high-electron-mobility transistor (HEMT) amplifier is used as the second stage amplifier at 4.5~K and its output is further amplified with a low-noise warm amplifier (WAMP) at room temperature. We use a shot noise tunnel junction (SNTJ) to generate a known white noise power at the ORCK input. Our noise model not only accounts for the non-unity transmission efficiencies $\eta_k$ between each of the amplifier stages -- due to lossy microwave components and rf coaxial cables -- but also accounts for the ORCK's internal loss $\eta_{1\text{--}2}$. In our measurements $\eta_1$ is due only to components required for operation of the SNTJ: a bias tee, single-stage 4--12~GHz cryogenic isolator, and a 13~GHz lowpass filter. All these components have been cryogenically measured and their transmission efficiencies are know to within $\pm0.1$~dB. The isolator and lowpass filter are implemented to reduce uncertainty in the shot noise we present to the ORCK -- e.g. from the ORCK pump generating a spurious contribution to the SNTJ voltage.}
    \label{fig:noise_schematic}
\end{figure*}

To characterize the system noise with an ORCK as the FSA we use a shot-noise tunnel junction (SNTJ) to generate a known white noise power at the ORCK input in the experimental configuration, realized in a dilution refrigerator, shown in Fig.~\ref{fig:noise_schematic}. The room-temperature, or warm amplifier (WAMP), is used to boost the cryostat output signal above the detector noise of a VNA or spectrum analyzer. As the ORCK is a non-ideal, wideband parametric amplifier we necessarily adopt a noise model which encapsulates this behavior -- specifically the signal-idler gain asymmetry. Additionally, we take into account the non-negligible loss of interconnects and non-superconducting external rf components via their transmission efficiency at the $k^{\text{th}}$ amplification stage, $\eta_k$ \cite{malnou2024low}. We note that the SNTJ-ORCK transmission efficiency $\eta_1$ would normally encompass components required for connection and operation of both the SNTJ and the KIT, but in the case of the ORCK this is due only to the SNTJ-associated hardware and $\eta_1$ is correspondingly minimized when using an ORCK.

Formally, the output of the parametric FSA (the ORCK) at the signal frequency depends on the input at both the signal and idler frequencies, $N_{\text{in}}^{\{s, i\}}$, the signal-idler gain asymmetry $\tilde{G}_1^i / \tilde{G}_1^s$, and the ORCK excess noise $\tilde{N}_{\text{ex},1}^s$ -- which is the degree to which the parametric FSA fails to reach the QL. In photon-normalized units we have
\begin{equation}
    \tilde{N}_{\text{out},1}^s = \tilde{G}_1^s\left( \tilde{N}_{\text{in}}^s + \frac{\tilde{G}_1^i}{\tilde{G}_1^s} \tilde{N}_{\text{in}}^i + \tilde{N}_{\text{ex},1}^s \right).
\end{equation}
Here the \textit{true gain} $\tilde{G}_1$ is the on/off gain $G_1$ reduced by the ORCK's off-state transmission efficiency, $\eta_{1\text{--}2}$, (internal loss)
\begin{equation}
    \tilde{G}_1 = G_1 \eta_{1\text{--}2}.
\end{equation}
Here and below we drop the signal-idler distinction when all terms in an expression are to be evaluated at only the signal or idler frequency. The actual input noise and excess noise are modified by $\eta_1$, which is modeled as a beamsplitter:
\begin{equation}
    \tilde{N}_{\text{in}} = N_{\text{in}} \eta_1 + N_{T_1}(1 - \eta_1)
\end{equation}
\begin{equation}
    \tilde{N}_{\text{ex},1}^s = \frac{N_{\text{ex},1}^s + N_{T_1}^s(1 - \eta_1^s)}{\eta_1^s} + \frac{\tilde{G_1^i}}{\tilde{G_1^s}} \frac{N_{1, ex}^i + N_{T_1}^i (1 - \eta_1^i)}{\eta_1^i}.
    \label{eq:N1ex}
\end{equation}
$N_{T_1}^{\{s, i\}}$ is the photon thermal occupancy at temperature $T_1$ at the signal and idler frequencies.

Since $\tilde{G}_2 \tilde{G}_1 > 50$~dB raises even the QL well above $N_{T_{300K}}$ we may neglect the added noise $N_{\text{add},3}$ and the loss between the second and third amplifier stages ($\eta_3 = 1$). The signal at the spectrum analyzer input is then
\begin{equation}
    N_{\text{out},3} = \tilde{G}_3 \tilde{G}_2 (\tilde{N}_{\text{out},1} + \tilde{N}_{\text{add},2})
\end{equation}
\begin{equation}
    N_{\text{out},3}^s = \tilde{G}_3^s \tilde{G}_2^s \tilde{G}_1^s \left( \tilde{N}_{\text{in}}^s + \frac{\tilde{G}_1^i}{\tilde{G}_1^s} \tilde{N}_{\text{in}}^i + \tilde{N}_{\text{ex},1}^s \right) + \tilde{G}_3^s \tilde{G}_2^s N_{\text{add},2}^s,
    \label{eq:Nout3_full}
\end{equation}
where $\tilde{N}_{\text{add},2}$ is the HEMT-added noise. In the high-ORCK-gain and low-HEMT-noise limit, i.e. when the ORCK gain raises the 0.5 quanta QL well above the HEMT-added noise of $\lesssim 10$~quanta (when $G_1 \gtrsim 16$~dB) we may neglect the final term in Eq.~(\ref{eq:Nout3_full}). In this case the system noise becomes the QL plus the ORCK excess noise, as these are the two terms appearing in the readout chain output which are not due to the input signal.

Finally, it should be noted that the QL arises from the vacuum occupancy of the idler mode since this is always mixed into the output at the signal frequency via the signal-idler gain asymmetry -- even in the case the parametric amplifier is ideal (or very narrowband) with $G_1^i / G_1^s = 1$. In our formulation $\tilde{N}_{\text{ex},1}^s$ is actually the \textit{excess noise} not the system noise. Furthermore, in the case of nonzero excess idler noise and when the gain asymmetry is significant, which would occur at the edges of the gain bandwidth, Eq.~(\ref{eq:N1ex}) suggests higher system noise when $\omega^s > \omega^i$.

\section{Design of On-chip rf Components}
To maintain flexibility we choose to implement wideband bias tees and directional couplers in the first ORCK design. With this choice we can freely tune future ORCK operating frequencies while requiring minimal or no changes to the integrated bias circuits -- not the case for a diplexer-based implementation which must be redesigned whenever the ORCK operating frequency is changed.

\begin{figure}
    \centering
    \includegraphics[width = .49\textwidth]{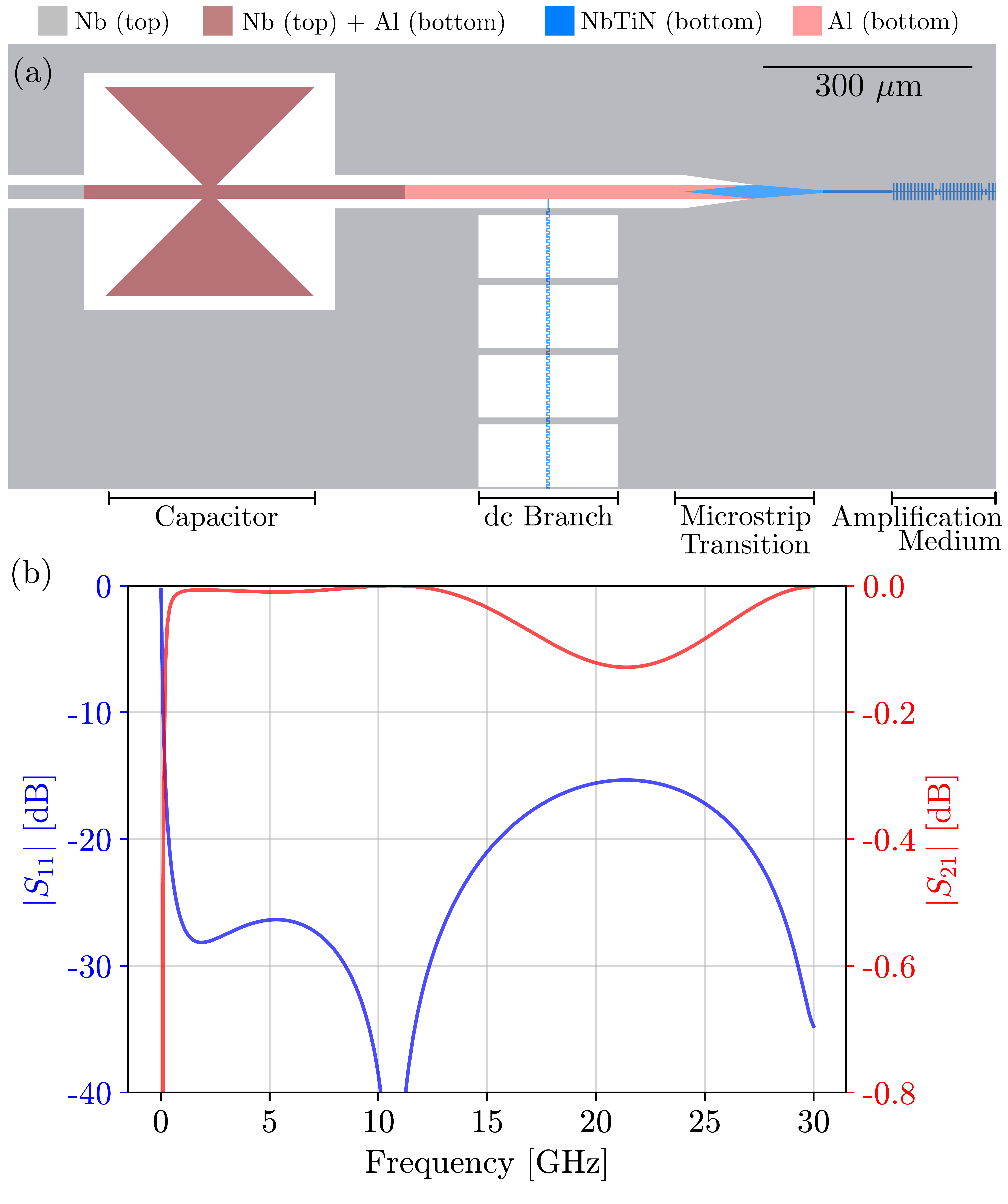}
    \caption{\textbf{(a)} Layout of the ORCK bias tee. The dc block is formed from the bowtie overlap capacitor and a 100~nm $\alpha$-Si dielectric. The dc branch is formed from the 2~$\mu$m wide meandered NbTiN connecting to the CPW from the bottom. Ground plane reliefs further increase the characteristic impedance of the dc line and we leave crossovers to improve rf grounding at chip scale. Also shown is the CPW-inverted-microstrip transition. Gradually tapering in the high-$L_k$ NbTiN is relevant because the sheet inductance increases by over a factor of 50 when the NbTiN is introduced. To minimize the impedance mismatch throughout this section the capacitance is also increased as the NbTiN width is tapered in. \textbf{(b)} Simulation of the dc block and CPW-microstrip transition (no dc bias line) showing reflections are effectively reduced below -15~dB at frequencies as high as 30~GHz. The increase in $|S_{11}|$ between 15~GHz and 25~GHz is predominantly due to the length of the simulated structure and should not be present in fabricated devices.}
    \label{fig:dc_block}
\end{figure}

\subsection{Bias Tee}
The dc block portion of the ORCK bias tee is based on \cite{elsbury2010broadband} and utilizes bowtie overlap capacitor pads to maintain a short electrical length as high as 30~GHz -- while preserving a lower cutoff frequency of $\sim 50$~MHz. Fig.~\ref{fig:dc_block} shows the bias tee layout and electromagnetic simulation results (omitting the bias line).

Rather than implementing a multistage washer inductor network with shunt resistors to de-Q the inductor self-resonances \cite{elsbury2010broadband}, we choose instead to construct the dc branch of the bias tee by leveraging the large NbTiN kinetic (sheet) inductance intrinsic to the fabrication of the amplification medium. Lossy inductor shunts in the dc branch introduce insertion loss in the high-pass branch so our approach aims to mitigate this effect by eliminating the shunts. To achieve a low cutoff frequency -- below $\sim 100$~MHz ideally -- we require around 100~nH of total inductance on the dc line. With the 35~pH/$\square$ kinetic inductance native to the amplification medium this means we need 4000 squares to meet the 100~nH target; which is achievable with a small meander on the 10~mm ORCK die. Fig.~\ref{fig:dc_block}(a) shows the layout of the dc line where we apply reliefs in the ground plane with crossover-like bridges to create a large rf impedance ($\gtrsim 700~\Omega$) while preserving quality grounding over the whole chip.

\subsection{Directional Coupler}
\begin{figure}
    \centering
    \includegraphics[width = .49\textwidth]{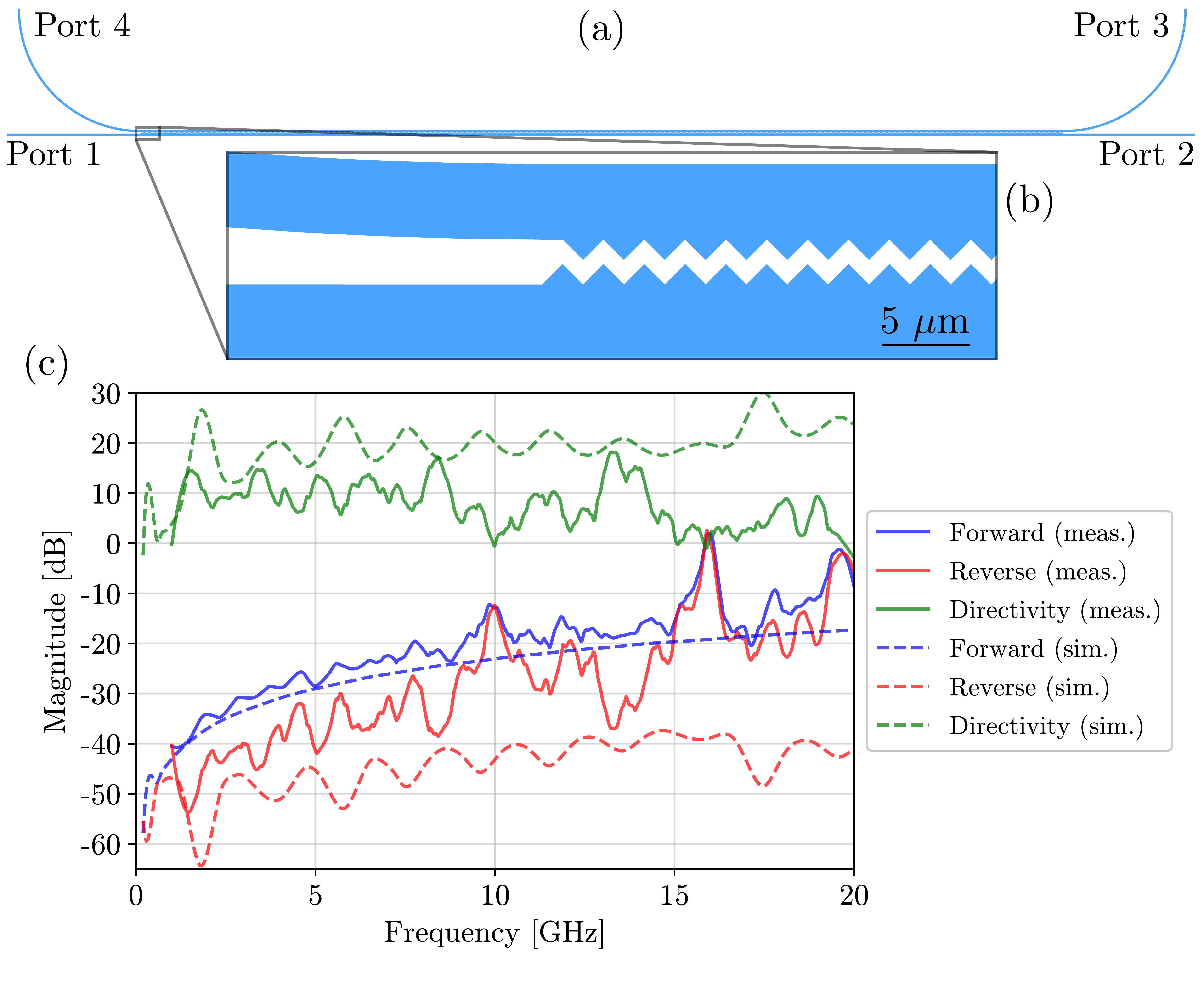}
    \caption{\textbf{(a)} Layout of the Podell wiggly coupler used for the ORCK where we have hidden the top ground plane layer to show the bottom layer traces only (NbTiN). The coupler is constructed of a zig-zag coupling region with a minimum line separation of 1~$\mu$m and total length of 1500~$\mu$m. $S_{21}$ describes the typical \textit{through} transmission and $S_{24}~(S_{23})$ describe the forward (reverse) coupling from the weakly coupled lines. The directivity is thus the ratio of the forward and reverse coupling: $S_{24} - S_{23}$ (in dB). \textbf{(b)} Zoom in of the first $\sim 25~\mu$m of the coupling region. \textbf{(c)} Simulated and measured performance of the ORCK coupler. Measurements were performed at 3~K using a test chip omitting the amplification medium. Data has been corrected via a calibration to remove the cryostat's ingress/egress cabling loss.}
    \label{fig:wc_gds_sim_and_meas}
\end{figure}

A directional coupler which is wideband, has high directivity, and can be adapted to different frequencies is desirable. To this end we choose the Podell wiggly coupler created from a sawtooth-shaped gap in the coupling region \cite{podell1970high, li2022wideband, shauerman2011development}. In our fabrication process the ground plane spacing of 100~nm, combined with the minimum (optical) lithographical feature size of 1~$\mu$m in the amplification medium, presents challenges in creating compact coupler for the ORCK. Our target forward coupling of $-20$~dB~$\pm~5$~dB in the 10~GHz--20~GHz frequency range necessitates increasing the length of the coupler to 1500~$\mu$m. Fortunately there is ample space for this structure on our 10~mm~$\times$~10~mm dies. Fig.~\ref{fig:wc_gds_sim_and_meas}(a) and (b) show the layout of the directional coupler, while dashed lines in Fig.~\ref{fig:wc_gds_sim_and_meas}(c) show the simulated forward and reverse coupling, and the directivity.

\section{Experimental Results}

\subsection{Bias Tee and Directional Coupler Cryogenic Performance}
We fabricate rf structure test chips on which we replace the long amplification medium with a minimal-length 50~$\Omega$ inverted microstrip line. Otherwise these chips are identical to the ORCK. We perform $S$-parameter measurements cryogenically of the bias tees and coupler that are forward/reverse-transmission-calibrated to remove ingress/egress cabling loss. 

Fig.~\ref{fig:wc_gds_sim_and_meas}(c) shows the behavior of the directional coupler on the rf test chip measured at 3~K. The forward coupling is close to the simulation result, while the directivity is approximately 10~dB lower than expected. In practice the lower-than-expected directivity is only of concern in the event the DUT being read out by the ORCK requires significant protection from a backward-traveling pump mode. This is because the traveling-wave nature of the ORCK makes its performance resilient to any reverse-coupled pump power -- since this power is both counter-propagating to the signal (readout) tones, and does not travel through the amplification medium.

Unfortunately we find that the dc line of the bias tee acts as a very high-$Q$ resonator at 900~MHz that is capable of ringing at all harmonics up to (and likely above) 20~GHz. This behavior was not seen during the design phase as simulating the entire bias tee structure, with a characteristic maximum (minimum) length scale of $5000~\mu$m ($0.5~\mu$m), is intractable. However, in Fig.~\ref{fig:orck_gain}(b) we do demonstrate a low frequency cutoff below 1~GHz for both the dc block (highpass cutoff) and the dc bias line (lowpass cutoff) which validates the primary functionality of the on-chip bias tee for ORCK. The observed dc bias line resonance can be eliminated in future ORCKs via implementation of de-Q-ing techniques \cite{elsbury2010broadband} (with attention to minimize the insertion loss), or substitution of the bias line with a lowpass filter or impedance matching network \cite{kaufman2023josephson, matthaei1980microwave}.

\subsection{ORCK Gain}
\begin{figure}
    \centering
    \includegraphics[width = .49 \textwidth]{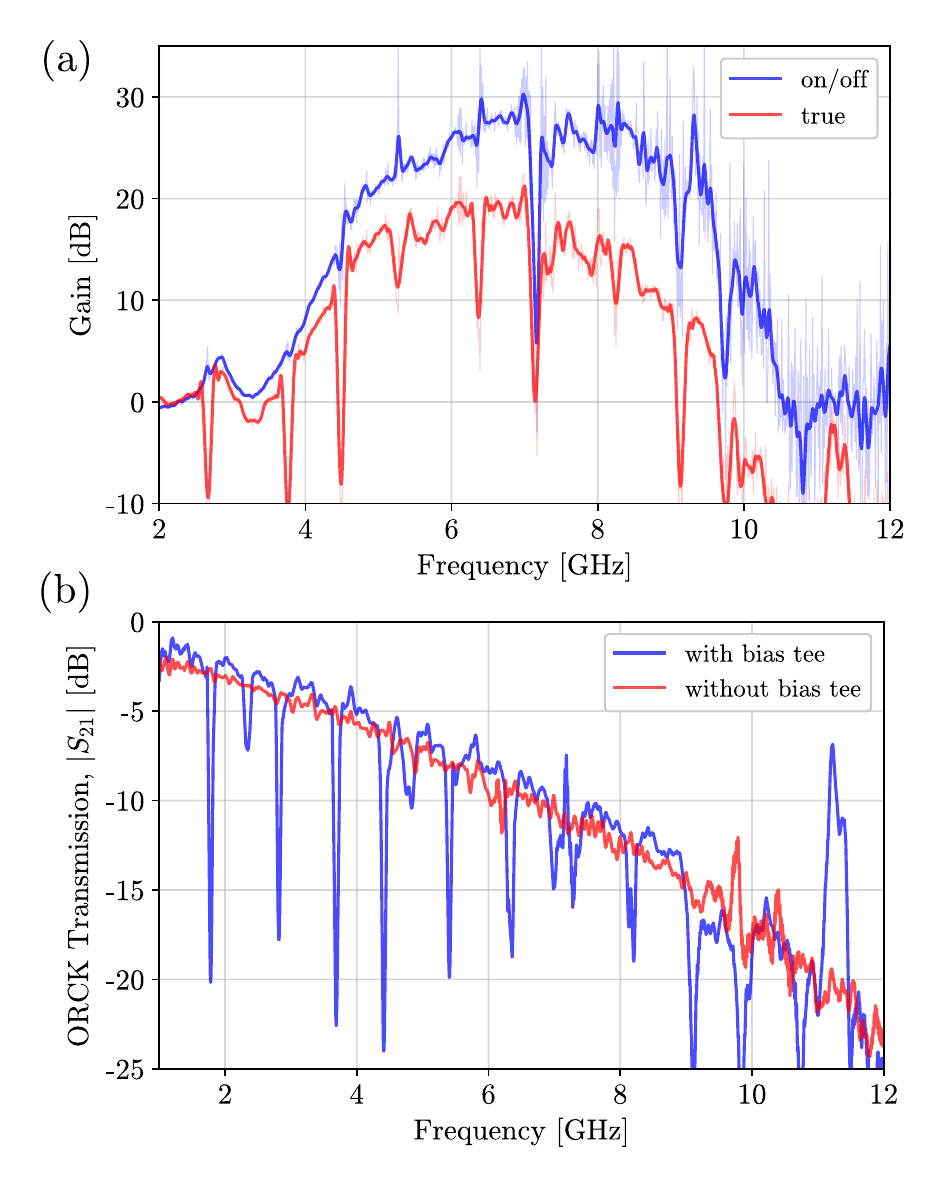}
    \caption{\textbf{(a)} ORCK gain measured at 50~mK with the raw data shown in light colors and a darker, smoothed trace to guide the eye. Large dips in the gain curves are the result of the bias tee dc branch's resonances, which are clearly shown in (b). We report both the on/off and the true gain. In our case up to a 15~dB discrepancy in the on/off and true gains is observed; underscoring the importance of this distinction. \textbf{(b)} ORCK transmission (internal loss) measurement, After ingress/egress cable loss calibration. We show data for both a 6-port ORCK, i.e. including the bias tees and directional coupler, and for a 4-port ORCK where the bias tees are omitted. Here we clearly demonstrate the unwanted resonances arise from addition of the bias tee as it is currently designed.}
    \label{fig:orck_gain}
\end{figure}

Device screening is typically performed in a 3~K cryostat to identify functional and high-performing devices, after which these devices are further characterized inside a dilution refigerator at 50~mK. Fig.~\ref{fig:orck_gain}(a) shows the post-optimization gain curve of a representative ORCK (at 50~mK) with both the \textit{on/off} and the \textit{true} gain. The on/off gain is simply the ratio of $|S_{21}|$ with the pump on to the pump off. The true gain is the KIT's actual contribution to the system gain and is in fact the on/off gain reduced by the amplifier's internal loss. For most TWPAs (Josephson- and kinetic-inductance-based) this loss is non-negligible and the discrepancy between the on/off and true gains can be significant -- making this an important distinction. Fig.~\ref{fig:orck_gain}(a) demonstrates that the on/off gain can be an over-estimate of the ORCK's true gain by 10~dB or more depedning on frequency. To measure the true gain we use two cryogenic switches to toggle between a through line (SMA female-female barrel connector) and the ORCK. Thus we measure the ingress/egress cabling loss combined with the HEMT and WAMP gain along the same path used to characterize the ORCK. Identical cabling lengths are used to connect the ORCK to the switches and to connect the SMA barrel.

\subsection{System Noise}
To perform a noise measurement using the setup shown in Fig.~\ref{fig:noise_schematic} we capture the readout chain output power at frequency $\omega^s$ on a spectrum analyzer while changing the ORCK input noise via controlling the voltage across the SNTJ. The asymptotic output of the SNTJ is, for $\hbar \omega^s \gg k_B T_1$, $e |V_{SNTJ}| / 2 h$ -- where $e$ is the elementary charge and $h$ Planck's constant \cite{malnou2024low}. The SA is operated in zero-span mode, with a  resolution bandwidth of 1~MHz and a video bandwidth of 8~MHz. The ORCK input power at the signal and idler frequencies is assumed to be equal because the frequency dependence of loss and reflections in the SNTJ package and interconnects is approximately negligible between 4~GHz and 10~GHz. We then fit the $\tilde{N}_{\text{out},3}^s$ vs $\tilde{N}_{\text{in}}^s = \tilde{N}_{\text{in}}^i = \tilde{N}_{\text{in}}$ data to the form
\begin{equation}
    \tilde{N}_{\text{out},3} = \tilde{G}_\text{sys} \left( \tilde{N}_{\text{in}}^s + \frac{\tilde{G}_1^i}{\tilde{G}_1^s} \tilde{N}_{\text{in}}^i + \tilde{N}_{\text{ex},1}^s \right)
    \label{eq:Nout3_simplified}
\end{equation}
with the system gain $\tilde{G}_\text{sys}^s = \tilde{G}_3^s \tilde{G}_2^s \tilde{G}_1^s$, signal-idler gain asymmetry $\tilde{G}_1^i / \tilde{G_1}^s$, and ORCK excess noise $\tilde{N}_{\text{ex},1}^s$ as free parameters. This is the high-ORCK-gain and low-HEMT-noise simplifcation of Eq.~\ref{eq:Nout3_full}.

\begin{figure}
    \centering
    \includegraphics[width=0.49\textwidth]{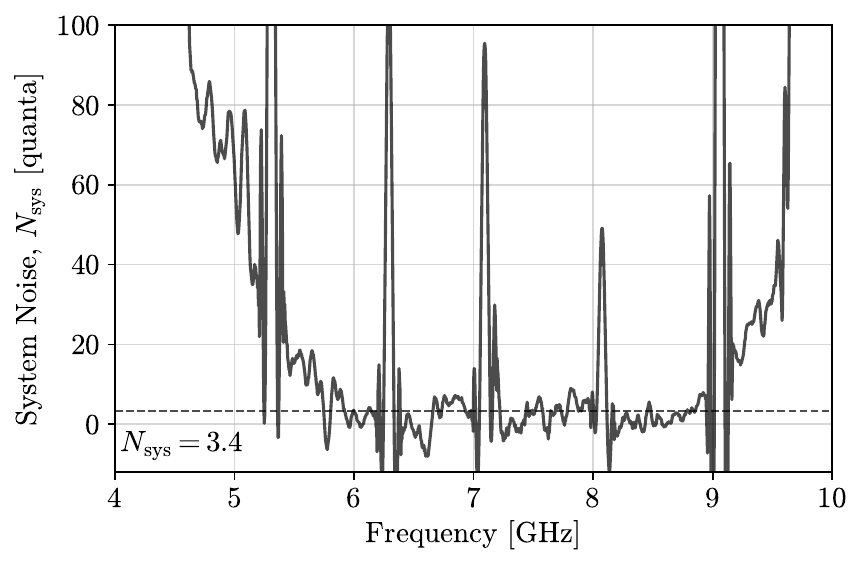}
    \caption{Extracted system noise with the ORCK as the FSA in the readout chain described by Fig.~\ref{fig:noise_schematic}. We use the noise model presented in Sec.~\ref{sec:noise_model} and fit to Eq.~(\ref{eq:Nout3_simplified}) to estimate the ORCK excess noise $N_{\text{ex},1}$. Note, because for the ORCK the bias tees and directional coupler are now constructed from low-loss superconducting circuits located on the amplifier die itself, we are free to move the reference plane directly to the ORCK input by setting $\eta_1 = 1$. Thus the modified excess noise $\tilde{N}_{\text{ex},1}$ becomes the true system excess noise $N_{\text{ex},1}$ with the reference plane at the ORCK package input. The dashed line represents the median system noise in the 3~dB bandwidth.}
    \label{fig:orck_noise}
\end{figure}

Note that naively leaving the bounds on the fitted parameters $\tilde{G}_\text{sys}$, $\tilde{G}_1^i / \tilde{G}_1^s$, and $\tilde{N}_{\text{ex},1}^s$ completely unconstrained, i.e. $[-\infty,~\infty]$, invariably results in nonphysical fits. Specifically we find the gain asymmetry tends to a local minimum near -0.995. This drives the extracted system gain far above its physical value and the excess noise almost identically to 0 over a bandwidth well in excess of the ORCK's 3~dB bandwidth. Imposing physical bounds, most notably $\tilde{G}_1^i / \tilde{G}_1^s \in [0.01,~100]$, yields physical results and we obtain the estimate of the system noise with the ORCK as the FSA shown in Fig.~\ref{fig:orck_noise}. 

In the high-ORCK-gain and low-HEMT-noise limit the total system noise is simply $N_{\text{sys}} = N_{\text{ex},1} + 0.5$ -- which is approximately valid for our device over its 3~dB bandwidth (see Fig.~\ref{fig:orck_gain}(a)). An important distinction for the ORCK is that this system noise is the true system noise, i.e. it is derived from $N_{\text{ex},1}$ and not $\tilde{N}_{\text{ex},1}$, because $\eta_1$ no longer plays any role in defining the ORCK's input reference plane. Instead $\eta_1$ is now solely affected by hardware choices necessitated by SNTJ operation or for maximizing the performance of a DUT replacing the SNTJ \cite{zhang2023traveling}. This means with the ORCK as the FSA, we demonstrate a 3~dB bandwidth median \textit{total} system noise under 3.5~quanta \textit{with the reference plane directly at the DUT output} -- which is now effectively the ORCK input. Contemporary JTWPAs demonstrate system noise performance either similar \cite{malnou2024low} or multiple times higher \cite{qiu2023broadband, ranadive2021reversed} than the ORCK and often at reference planes upstream of lossy components that degrade the system noise when referred to the DUT input.

\section{Conclusion}
Here we have demonstrated a significant advancement in traveling wave parametric amplifier (TWPA) design by integrating all mandatory rf bias circuitry on-chip in a kinetic inductance TWPA (KIT) -- which we call the on-chip rf component KIT (ORCK). The ORCK footprint is reduced by more than five times via integration of two on-chip bias tees (for dc bias application) and a directional coupler (for pump injection). The directional coupler, based on a Podell wiggly coupler \cite{podell1970high}, supplies the desired coupling of $\sim-20$~dB between 10~GHz and 20~GHz and agrees well with simulation. The bias tee high pass and low pass branches both achieve cutoff frequencies well below 1~GHz and, in conjunction with the on-chip directional coupler, provide the first-ever demonstration of a KIT requiring no external rf components (diplexers, bias tees, directional couplers, etc.) to operate. Resonances in the bias line will be eliminated in the future via redesign of the bias tee dc branch using filter synthesis or de-Q-ing techniques to recover a cleaner gain profile.

Furthermore, the ORCK presented herein displays a 2.8~GHz 3~dB bandwidth with a median \textit{true gain} of 17.5~dB, and a median total system noise below 3.5~quanta. This noise performance is competitive with contemporary JTWPAs \cite{malnou2024low, qiu2023broadband, ranadive2021reversed} and lower than prior KIT demonstrations \cite{malnou2021three}. Of course the primary advantages the ORCK grants an experimenter are the significant reduction in complexity and the fact that the parametric amplifier input reference plane is now truly the ORCK package input -- now much closer to the DUT output reference plane. This is opposed to the case where the reference plane is at the input of a chain of one or more external components (bias tees, directional couplers, or diplexers); as in conventional parametric amplifiers lacking on-chip component integration. Substitution of these lossy commercial components with the ORCK's on-chip superconducting counterparts grants an additional improvement in system noise -- in some cases approaching a 2 quanta reduction \cite{malnou2024traveling}. Future highly sensitive physics, astronomy, and quantum information experiments utilizing an ORCK as the readout chain FSA will benefit significantly from its drastic miniaturization and enhanced noise performance.

\section*{Acknowledgment}
This work is supported by the National Aeronautics and Space Administration (NASA) under Grant No. NNH18ZDA001N-APRA, the Department of Energy (DOE) Accelerator and Detector Research Program under Grant No. 89243020SSC000058, and DARTWARS, a project funded by the European Union’s H2020-MSCA under Grant No. 101027746. The work is also supported by the Italian National Quantum Science and Technology Institute through the PNRR MUR Project under Grant PE0000023-NQSTI.

\bibliographystyle{ieeetr}
\bibliography{references.bib}



\end{document}